\documentclass[12pt]{article}
\usepackage[dvips]{graphicx,psfrag}
\usepackage{amssymb}
\usepackage{color}
\input epsf
\definecolor{Blue}{rgb}{0.3,0.3,0.9}
\definecolor{Red}{rgb}{1,0,0}
\definecolor{Green}{rgb}{0,1,0}
\newcommand{\be}{\begin{equation}}
\newcommand{\ee}{\end{equation}}
\newcommand{\bea}{\begin{eqnarray}}
\newcommand{\eea}{\end{eqnarray}}

\begin{document}
\begin{titlepage}
\begin{center}
{\bf GRAVITINO PRODUCTION BY PRIMORDIAL BLACK HOLE EVAPORATION AND CONSTRAINTS ON
THE INHOMOGENEITY OF THE EARLY UNIVERSE}
\vskip 2 cm
{\bf M. Yu. Khlopov}
\vskip 0.4cm
 Physics Department,
Rome University 1 "La Sapienza"\\
P. le A. Moro, 2, 00185, Rome, Italy\\
Cosmion, Keldysh Institute of Applied Mathematics\\ 
Miusskaya pl. 4, 125047, Moscow, Russia \\
Moscow Engineering Physics Institute\\
Kashirskoe shosse 31, 115409, Moscow, Russia
\vskip 0.7cm
{\bf A. Barrau, J. Grain}
\vskip 0.4cm
Laboratory for Subatomic Physics and Cosmology\\
 Joseph Fourier University, CNRS-IN2P3 \\
 53, avenue des Martyrs, 38026 Grenoble cedex, France

\end{center}
\date{\today}

\vskip 1cm
\small
\begin{center}
{\bf Abstract}
\end{center}

\begin{quote}
In supergravity models, the evaporation of light Primordial Black Holes (PBHs) should be a source of 
gravitinos. By considering this process, new stringent
limits are derived on the 
abundance of small black holes with initial masses less than $10^9$~g. 
In minimal 
supergravity, the subsequent decay of evaporated gravitinos into cascades of non-equilibrium particles 
leads to the formation of elements whose abundance is constrained by observations. 
In gauge mediated supersymmetry breaking models, their density is required not 
to overclose the Universe. As a result, cosmological models with substantial 
inhomogeneities on small scales are excluded.\end{quote} 
\normalsize
\date{\today}

\end{titlepage}

\section{Introduction}

Primordial Black Holes (PBHs) are a very sensitive cosmological probe for physics
phenomena occurring in the early Universe. They could be formed by many different
mechanisms, {\it e.g.}, initial density inhomogeneities \cite{hawking1}, a softening of the
equation of state \cite{canuto,khlopov0, khlopov1}, collapse of cosmic strings \cite{hawking2}, a
double inflation scenario \cite{nas}, phase transitions \cite{hawking3}, a
step in the power spectrum \cite{polarski1}, etc. (see
\cite{khlopov2} for a review). Constraining the abundance of primordial black
holes can lead to invaluable information on cosmological processes, particularly
as they are probably the only viable probe for the power
spectrum on very small scales which remain far from the Cosmological Microwave
Background (CMB) and Large Scale Structures (LSS) sensitivity ranges. To date, 
only PBHs with initial masses between $\sim 10^9$~g and $\sim 10^{16}$~g
have led to stringent limits (see {\it e.g.} \cite{carr1,polnarev}) from consideration of
the entropy per baryon,
the deuterium destruction, the $^4$He destruction and the cosmic-rays currently
emitted by the Hawking process \cite{hawking4}. The existence of light 
PBHs should lead to important observable constraints, either through the
direct effects of the evaporated particles (for initial masses between $10^{14}$~g and 
$10^{16}$~g) or through the indirect effects of their interaction with matter and 
radiation in the early Universe (for PBH masses between $10^{9}$~g and 
$10^{14}$~g). In these constraints, the effects taken into 
account are those related with known particles. However, since the evaporation products 
are
created by the gravitational field, any quantum with a mass lower than the black
hole temperature should be emitted, independently of the strength of its
interaction. This could provide a copious production of superweakly interacting
particles that cannot not be in equilibrium with the hot plasma of the very early Universe.

Following \cite{khlopov2,khlopov6,khlopov7} and \cite{lemoine,green1} (but in a
different framework and using more stringent constraints), 
this article derives new limits on the mass fraction of black
holes at the time of their formation ($\beta \equiv \rho_{PBH}/\rho_{tot}$) using the production of gravitinos
during the evaporation process. Depending on whether gravitinos are expected to
be stable or metastable, the limits are obtained using the requirement that they
do not overclose the Universe and that the formation of light nuclei
by the interactions of $^4$He nuclei with nonequilibrium
flux of D,T,$^3$He and $^4$He does not contradict the observations. This approach
is more constraining than the usual study of photo-dissociation induced 
by photons-photinos pairs emitted by decaying gravitinos. It opens a
new window for the upper limits on $\beta$ below $10^9$~g. The cosmological
consequences of our new limits are briefly reviewed in the framework
of 3 different scenarios~: a
blue power spectrum, a step in the power spectrum and first order phase
transitions.

\section{New limits on the PBH density}

Several constraints on the density of PBHs have been derived in different
mass ranges assuming the evaporation of only standard model particles~: for $10^9~{\rm g}<M<10^{13}~{\rm g}$ the entropy per baryon at
nucleosynthesis  was used \cite{mujana} to obtain $\beta < (10^9~{\rm g}/M)$, for 
$10^9~{\rm g}<M<10^{11}~{\rm g}$ the production of $n\bar{n}$ pairs at nucleosynthesis was
used \cite{zeldovich} to obtain $\beta  < 3\times 10^{-17}
(10^9~{\rm g}/M)^{1/2}$ , for $10^{10}~{\rm g}<M<10^{11}~{\rm g}$ deuterium destruction
was used \cite{lindley} to obtain $\beta  < 3\times 10^{-22}
(M/10^{10}~{\rm g})^{1/2}$, for $10^{11}~{\rm g}<M<10^{13}~{\rm g}$ spallation
of $^4$He was
used \cite{vainer,khlopov6} to obtain $\beta  < 3\times 10^{-21}
(M/10^9~{\rm g})^{5/2}$, for $M\approx 5\times 10^{14}~{\rm g}$ the gamma-rays and
cosmic-rays were used \cite{macgibbon,barrau} to obtain $\beta <
10^{-28}$. Slightly more stringent limits where obtained in \cite{kohri}, leading
to $\beta < 10^{-20}$ for masses between $10^{9}~{\rm g}$ and $10^{10}~{\rm g}$ and in
\cite{barraugamma}, leading to $\beta < 10^{-28}$ for $M=5\times 10^{11}~{\rm g}$.
Gamma-rays and antiprotons where also recently re-analyzed in \cite{barraupbar} and 
\cite{custodio}, improving a little the previous estimates.
Such constraints, 
related to phenomena occurring after the nucleosynthesis, apply only for
black holes with initial masses above $\sim 10^9$~g. Below this value, the only limits
are the very weak entropy constraint (related with the photon-to-baryon ratio) and 
the quite doubtful relics constraint (assuming stable black holes 
forming at the 
end of the evaporation mechanism as described, {\it e.g.}, in \cite{alexeyev}).\\

To derive a new limit in the initial mass range $M_{Pl}<M<10^{11}$~g, gravitinos emitted
by black holes are considered in this work. Gravitinos are expected to be
present in all local supersymmetric models, which are regarded as the more
natural extensions of the standard model of high energy physics (see, {\it
e.g.}, \cite{olive} for an introductory review). 
In the framework of minimal Supergravity (mSUGRA), 
the gravitino mass is, by construction, expected to lie around the electroweak scale, {\it i.e.} in the
100 GeV range. In this case, the gravitino is {\it metastable} and decays after
nucleosynthesis, leading to important modifications of the nucleosynthesis
paradigm. Instead of using the usual photon-photino decay channel, this study
relies on the more sensitive gluon-gluino channel. Based on \cite{khlopov3}, the
antiprotons produced by the fragmentation of gluons emitted by decaying 
gravitinos are considered as
a source of nonequilibrium light nuclei resulting from collisions of those
antiprotons on equilibrium nuclei. Then, $^6$Li, $^7$Li and $^7$Be nuclei 
production by the interactions of the nonequilibrium nuclear flux with $^4$He 
equilibrium nuclei is taken into account and compared with 
data (this approach is supported by several recent analysis \cite{Karsten} which lead to
similar results).
The resulting Monte-Carlo estimates  \cite{khlopov3} 
lead to the following constraint on the concentration of gravitinos: $n_{3/2}<
1.1\times 10^{-13}m_{3/2}^{-1/4}$ where $m_{3/2}$ is the gravitino mass in GeV.
This constraint has
been successfully used to derive an upper limit on the reheating temperature of
the order \cite{khlopov3}: $T_R < 3.8\times 10^6$~GeV. The consequences of
this limit on cosmic-rays emitted by PBHs was considered, 
{\it e.g.}, in \cite{barrauprd}.
In the present approach, we relate this stringent constraint on the gravitino
abundance to the density of PBHs through the direct gravitino
emission. The usual
Hawking formula \cite{hawking4} is used for the number of particles of type
$i$ emitted per unit of time $t$ and per unit of energy $Q$.
Introducing the  temperature defined by
$
T=hc^3/(16\pi k G M)\approx(10^{13}{\rm g})/{M}~{\rm GeV},
$
taking the relativistic approximation for $\Gamma_s$,
and integrating over time and energy, the total number of quanta of type $i$ can
be estimated as:
$$N_i^{TOT}=\frac{27\times 10^{24}}{64\pi ^3
\alpha_{SUGRA}}\int_{T_i}^{T_{Pl}}\frac{dT}{T^3}\int_{m/T}^x\frac{x^2dx}{e^x-(-1)^s}
$$
where $T$ is in GeV, $M_{Pl}\approx 10^{-5}$~g, $x\equiv Q/T$, $m$ is the particle mass and $\alpha_{SUGRA}$
accounts for the number of degrees of freedom through $M^2dM=-\alpha_{SUGRA}dt$
where $M$ is the black hole mass.
Once the PBH temperature is higher than the gravitino mass, gravitinos will
be emitted with a weight related with their number of degrees of freedom. 
Computing the number of emitted gravitinos as a
function of the PBH initial mass and matching it with the limit on the gravitino density
imposed by nonequilibrium nucleosynthesis of light elements leads to an upper
limit on the PBH number density. If PBHs are formed during a radiation
dominated stage, 
this limit can easily be converted into an upper limit
on $\beta$ by evaluating the energy density of the radiation at the formation
epoch. The resulting limit is shown on Fig. 1 and leads to an important
improvement over previous limits, nearly independently of the gravitino mass 
in the interesting range. This opens a new window on the very small scales in the
early Universe.\\

\begin{figure}
	\begin{center}
		\includegraphics[scale=0.7]{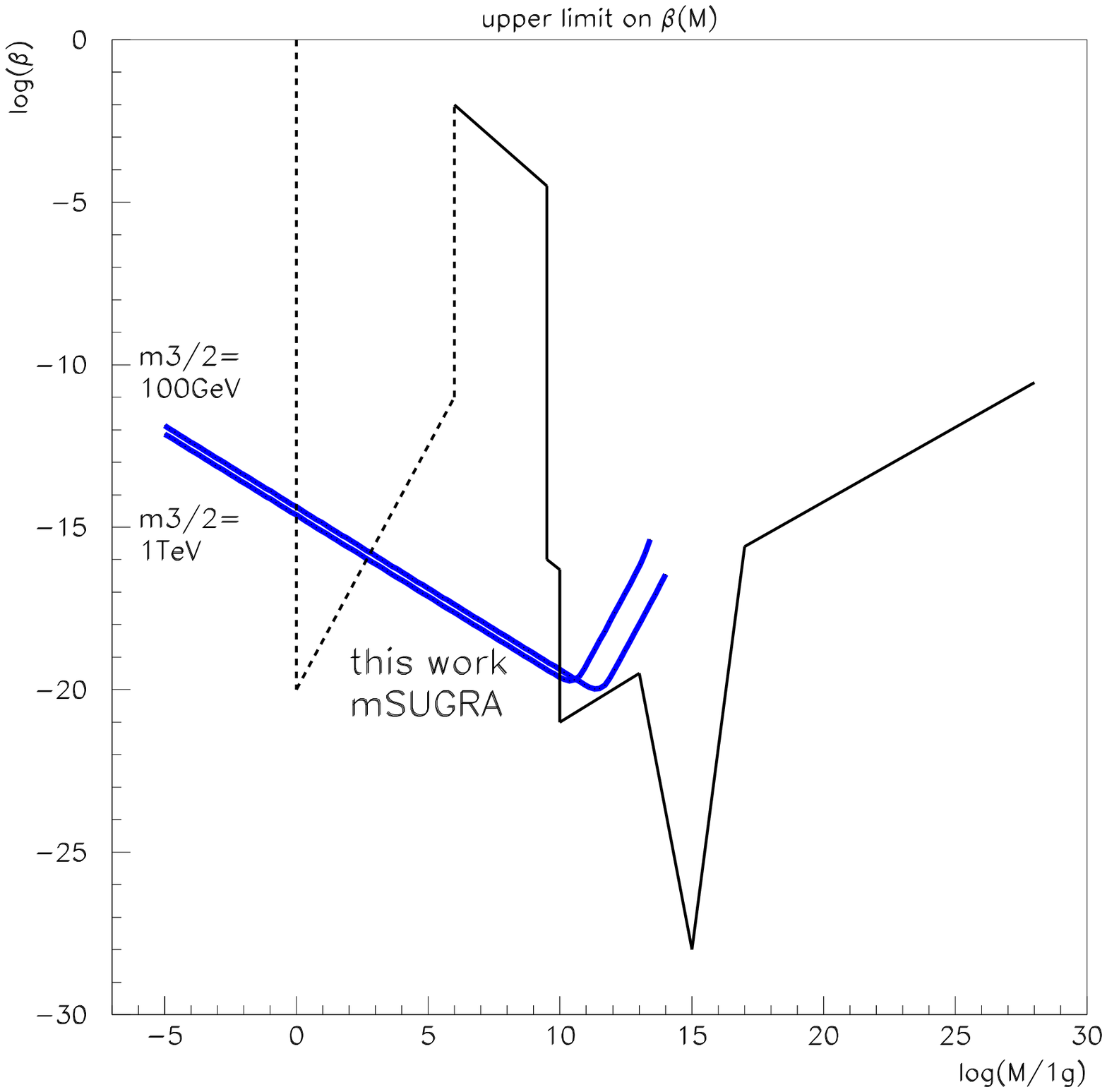}
		\caption{Constraints on the fraction of the Universe going into PBHs (adapted from
\cite{carr1,polnarev}). The two curves
obtained with gravitinos emission in mSUGRA correspond to $m_{3/2}$ = 100 GeV (lower curve in
the high mass range) and  $m_{3/2}$ = 1 TeV (upper curve in the high mass range)}
		\label{pot}
	\end{center}
\end{figure}

It is also possible to consider limits arising in  Gauge Mediated Susy
Breaking (GMSB) models \cite{kolda}. Those alternative scenarios, incorporating
a natural suppression of the rate of flavor-changing
neutral-current due to the low energy scale, predict the gravitino to be the
Lightest Supersymmetric Particle (LSP). The LSP is stable if R-parity is
conserved. In this case, the limit is obtained by
requiring $\Omega_{3/2,0}<\Omega_{M,0}$, {\it i.e.} by requiring that the
current gravitino density  does not exceed the matter density. It can
easily be derived from the previous method, by taking into account 
the dilution of gravitinos in the period of PBH evaporation and conservation of gravitino to specific entropy ratio, that~:
$$\beta \leq \frac{\Omega_{M,0}}{N_{3/2}\frac{m_{3/2}}{M}\left(
\frac{t_{eq}}{t_{f}} \right) ^{\frac{1}{2}}}$$
where $N_{3/2}$ is the total number of gravitinos emitted by a PBH with initial
mass $M$, $t_{eq}$ is the end of RD stage and $t_f=max(t_{form},t_{end})$ when a non-trivial equation of state for the period of PBH formation is considered, {\it e.g.} 
a dust-like phase which ends at $t_{end}$ \cite{khlopov8}.
The limit does not imply thermal equilibrium of relativistic plasma in the period before PBH evaporation and is valid even for low reheating temperatures provided that the equation of state on the preheating stage is close to relativistic.
With the present matter density $\Omega_{M,0}\approx 0.27$ \cite{wmap} 
this leads to the limits shown in Fig. 2 for two extreme cases: 
$m_{3/2}=10^{-5}$~GeV and $m_{3/2}=10$~GeV. These results are close to the
previous ones and remain very competitive in this mass range.
Models of gravitino dark matter with
$\Omega_{3/2,0} = \Omega_{CDM,0}$, corresponding to the case of equality in the
above formula,  were recently considered in \cite{Jedamzik1}.

\begin{figure}
	\begin{center}
		\includegraphics[scale=0.7]{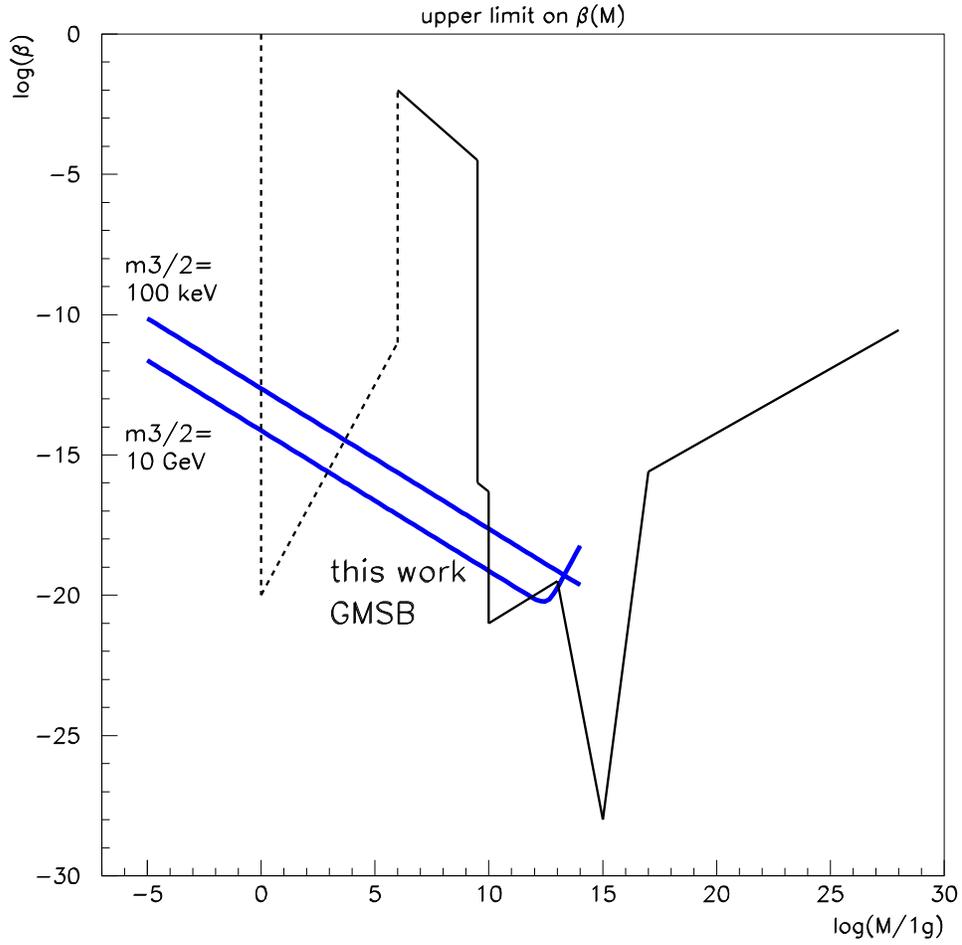}
		\caption{Constraints on the fraction of the Universe going into PBHs (adapted from
\cite{carr1,polnarev}). The two curves
obtained with gravitinos emission in GMSB correspond to $m_{3/2}=10^{-5}$ GeV (lower curve in
the high mass range) and  $m_{3/2}=10$ GeV (upper curve in the high mass range)}
		\label{pot}
	\end{center}
\end{figure}

\section{Cosmological consequences}

Our new stringent upper limits on the fraction of the Universe in primordial black holes
can be converted into cosmological constraints on models with significant power on small
scales.\\

The easiest way to illustrate the importance of such limits is to consider a blue power
spectrum and to derive a related upper value on the spectral index $n$ of scalar
fluctuations ($P(k)\propto k^n$). It has recently been shown by WMAP \cite{wmap} that
the spectrum is nearly of the Harrison-Zel'dovich type, {\it i.e.} scale invariant with
$n\approx 1$. However this measure was obtained for scales between $10^{45}$ and $10^{60}$ times
larger that those probed by PBHs and it remains very important to probe the power available
on small scales. The limit on $n$ given in this paper must therefore be understood as a way to constrain
$P(k)$ at small scales rather than a way to measure its derivative at large scales : it
is complementary to CMB measurements. Using
the usual relations between the mass variance at the PBH formation time
$\sigma_H(t_{form})$ 
and the same quantity today $\sigma_H(t_0)$ \cite{green},
$$\sigma_H(t_{form})=\sigma_H(t_0)\left(\frac{M_H(t_0)}{M_H(t_{eq})}\right)^{\frac{n-1}{6}}
\left(\frac{M_H(t_{eq})}{M_H(t_{form})}\right)^{\frac{n-1}{4}}$$
where $M_H(t)$ is the Hubble mass at time $t$ and $t_{eq}$ is the equilibrium time, it is
possible to set an upper value on $\beta$ which can be expressed as
$$\beta\approx \frac{\sigma_H(t_{form})}{\sqrt{2\pi}\delta_{min}}e^{-\frac{\delta_{min}^2}
{2\sigma_H^2(t_{form})}}$$
where $\delta_{min}\approx 0.3$ is the minimum density contrast required to form a PBH. The
limit derived in the previous section leads to $n<1.20$ in the mSUGRA case 
whereas the usually derived limits range between 1.23 and 1.31 \cite{green,kim}. 
In the GMSB case, it remains at the same level for $m_{3/2}\sim10$~GeV and is
slightly relaxed to $n<1.21$ for $m_{3/2}\sim100$~keV.
This substantial improvement is due to the much more important range of masses
probed and where derived for $M\sim 1$~g.\\

In the standard cosmological paradigm of inflation, the primordial power spectrum is
expected to be nearly --but not exactly-- scale invariant \cite{liddle}. The sign of the
running can, in principle, be either positive or negative.
It has been recently shown that models with a positive running
$\alpha_s$, defined as
$$P(k)=P(k_0)\left( \frac{k}{k_0} \right)^{n_s(k_0)+\frac{1}{2}\alpha_s ln \left(
\frac{k}{k_0}\right)},$$
are very promising in the framework of supergravity inflation (see, {\it e.g.},
\cite{kawa}). Our analysis strongly limits a positive running,
setting the upper bound at a tiny value $\alpha_s<2\times 10^{-3}$. This result is
more stringent than the upper limit obtained through a combined analysis of Ly$\alpha$
forest, SDSS and WMAP data \cite{seljak}, $-0.013<\alpha_s<0.007$, as it deals with scales
very far from those probed by usual cosmological observations. The order of magnitude of
the running naturally expected in most models --either inflationary ones (see,
{\it e.g.}, 
\cite{peiris}) or alternative ones (see, {\it e.g.}, \cite{khoury})-- being of a few times 
$10^{-3}$ our upper bound should help to distinguish between different scenarios.\\

In the case of an early dust-like stage in the cosmological evolution
\cite{khlopov0,polnarev,khlopov2}, the PBH formation probability is increased to
$\beta > \delta ^ {13/2}$
where $\delta$ is the density contrast for the considered small scales. The associated limit on $n$ is
strengthened to $n<1.19$.\\

Following \cite{green}, it is also interesting to consider primordial density
perturbation spectra with
both a tilt and a step. Such a feature can arise from underlying physical
processes \cite{starobinsky} and allows investigation of a wider class of inflaton
potentials. If the amplitude of the step is defined so that the power on small
scales is $p^{-2}$ times higher than the power on large scales, the maximum 
allowed value for the spectral index can be computed as a function of $p$. 
Figure~3 shows those limits, which become extremely stringent when $p$ is
small enough, for both the radiation-dominated and the
dust-like cases. The different values of the gravitino mass
considered in the first section of this paper are all included within the width of the 
lines.\\

\begin{figure}
	\begin{center}
		\includegraphics[scale=0.7]{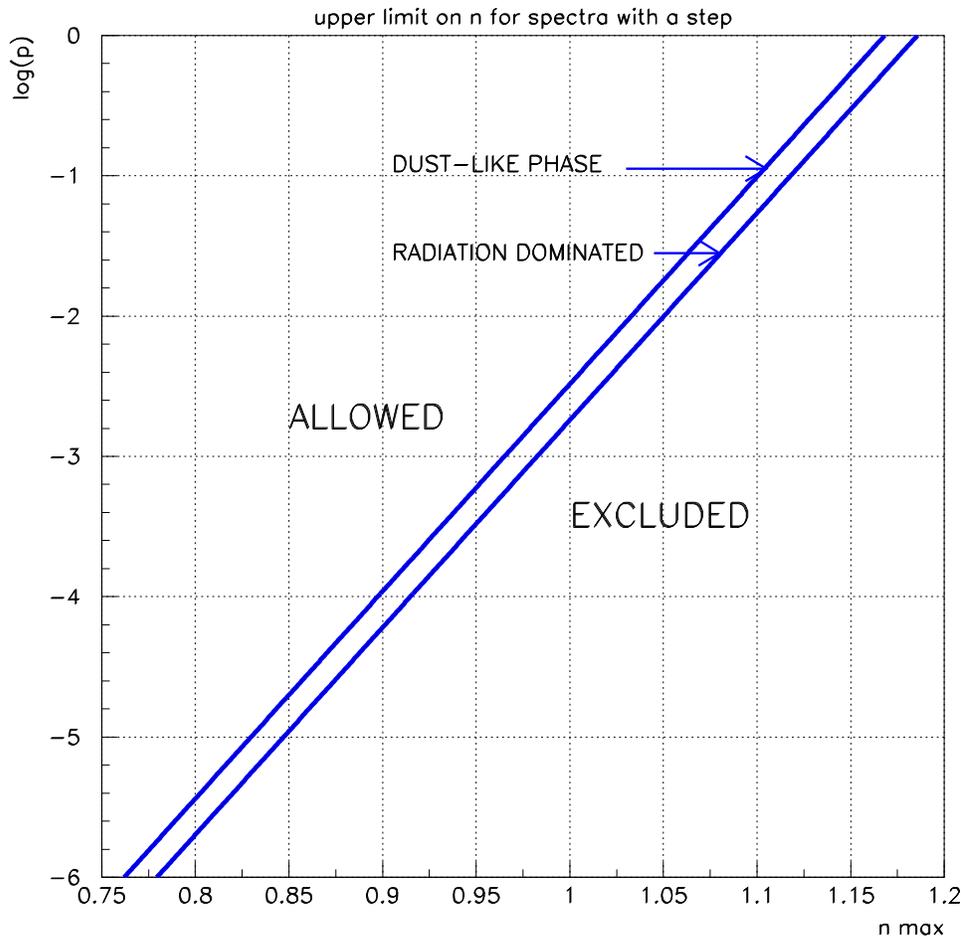}
		\caption{Upper limit on the spectral index of the power spectrum as a function
of the amplitude of the step.}
		\label{pot}
	\end{center}
\end{figure}

Another important consequence of our new limits concerns PBH relics
dark matter. The idea, introduced in \cite{macgibbon2}, that relics possibly
formed at the end of the evaporation process could account for the cold dark
matter has been extensively studied. The amplitude of the power boost required
on small scales has been derived, {\it e.g.}, in \cite{barrau2} as a function of the relic
mass and of the expected density. The main point was that the "step" (or
whatever structure in the power spectrum) should occur at low  masses to
avoid the constraints available between $10^9$~g and $10^{15}$~g. The limit on
$\beta$ derived in this work closes this dark matter issue except within a small
window below $10^3$~g.\\

Finally, the limits also completely exclude the possibility of a copious PBH formation
process in bubble wall collisions \cite{khlopov4}. This has important consequences
for the related constraints on first order phase transitions in the early Universe.

\section{Conclusion and prospects}

If local supersymmetry - supergravity - is the correct extension of the standard
model, the emission of
gravitinos from evaporating primordial black holes leads to important
constraints on their number density in the - so far nearly unexplored - low
formation mass
range. This allows us to exclude cosmological models with too much power on small
scales. In particular, a blue power spectrum or a positive running spectral
index are strongly disfavored. Any mechanism which would lead to small scale
inhomogeneities with a density contrast above $\sim 0.3$ beyond a tiny fraction
of the order of $10^{-12}-10^{-20}$ for masses between $10^{-5}$ and $10^{10}$g
are excluded by this analysis. This method offers nontrivial links between 
the inhomogeneity of the early Universe and the existence of (meta-)stable particles.\\

It should also be noticed that, as in inflationary cosmology,
the equilibrium is established only after reheating
at $T < T_R$. If $T_R$ is as low as \cite{khlopov3} $T_R < 3.8\times 10^6$~GeV,
superheavy particles with masses $m \gg T_R$ and superweakly interacting particles with
interaction cross section $\sigma \ll \frac{1}{T_R M_{Pl}}$ 
cannot be in equilibrium but can be copiously produced through
PBH evaporation. This property opens a wide range of possible applications
for testing particle theory and cosmological scenarios.

\section{Acknowledgment}

M.Kh is grateful to LPSC, CRTBT-CNRS and UJF, Grenoble, France for hospitality.

{}

\end{document}